\documentclass[preprint,showpacs,preprintnumbers,amsmath,amssymb]{revtex4}

\usepackage{graphicx}
\usepackage{dcolumn}
\usepackage{bm}

\newcommand{\qed}{\nobreak \ifvmode \relax \else
      \ifdim\lastskip<1.5em \hskip-\lastskip
      \hskip1.5em plus0em minus0.5em \fi \nobreak
      \vrule height0.75em width0.5em depth0.25em\fi}

\begin{document}

\preprint{}

\title{A Hypergeometric Formula Yielding Rational-Valued Hilbert-Schmidt Generic $2 \times 2$ {\it Generalized} Separability Probabilities}
\author{Paul B. Slater}%
\email{slater@kitp.ucsb.edu}
\affiliation{%
University of California, Santa Barbara, CA 93106-4030\\
}%
\date{\today}

\begin{abstract}
We significantly advance the research program initiated in "Moment-Based Evidence for Simple Rational-Valued Hilbert-Schmidt Generic $2 \times 2$ Separability Probabilities" ({\it J. Phys. A}, {\bf 45}, 095305 [2012]). A function $P(\alpha)$, incorporating a family of six hypergeometric functions, all with argument $\frac{27}{64} =(\frac{3}{4})^3$, is obtained here. It reproduces a series, $\alpha = \frac{1}{2}, 1, \frac{3}{2}, 2,\ldots,32$, of sixty-four Hilbert-Schmidt generic $2 \times 2$ {\it generalized} separability probabilities,  advanced on the basis of systematic high-accuracy probability-distribution-reconstruction computations, employing 7,501 determinantal moments of partially transposed $4 \times 4$ density matrices. For generic (9-dimensional) 
two-rebit systems, 
$P(\frac{1}{2}) = \frac{29}{64}$, (15-dimensional) two-qubit,  $P(1) = \frac{8}{33}$, and  (27-dimensional) two-quat(ernionic)bit systems,  
$P(2)=\frac{26}{323}$. The function $P(\alpha)$ is generated--applying the Mathematica command FindSequenceFunction--using solely {\it either} the thirty-two integral or half-integral probabilities, yet successfully predicts the unused 
thirty-two. 
\end{abstract}


\maketitle

\section{Introduction}

The predecessor paper \cite{MomentBased}--addressing the relatively 
long-standing $2 \times 2$ separability probability question \cite{ZHSL,
slaterJGP,slaterqip,slaterA,slaterC,slaterPRA,slaterPRA2,slaterJGP2,pbsCanosa,slater833} (cf. \cite{sz1,sz2,ye})--consisted largely of two sets of analyses. (In the foundational study \cite{ZHSL}, "three main reasons of importance"--philosophical, practical and physical--were given for studying this question.) The first set of analyses in \cite{MomentBased} was concerned with establishing formulas for the bivariate determinantal product moments 
$\left\langle \left\vert \rho^{PT}\right\vert ^{n}\left\vert \rho\right\vert
^{k}\right\rangle ,k,n=0,1,2,3,\ldots,$ with respect to Hilbert-Schmidt (Euclidean/flat) measure \cite{szHS} \cite[sec. 14.3]{ingemarkarol}, of generic (9-dimensional) two-rebit and (15-dimensional) two-qubit density matrices ($\rho$). Here 
$\rho^{PT}$ denotes the partial transpose of the $4 \times 4$ density matrix $\rho$. Nonnegativity of the determinant 
$|\rho^{PT}|$ is both a necessary and sufficient condition for separability in this $2 \times 2$ setting \cite{augusiak}.

In the second set of analyses in \cite{MomentBased}, the {\it univariate} determinantal moments $\left\langle \left\vert \rho^{PT}\right\vert ^{n} \right\rangle$ and $\left\langle \left ( \vert \rho^{PT}\right\vert \left\vert \rho\right\vert)^n
\right\rangle$, induced using the bivariate formulas, served as input to a Legendre-polynomial-based probability distribution reconstruction algorithm of Provost \cite[sec. 2]{Provost}. This yielded estimates of the desired separability probabilities. (The reconstructed probability distributions based on $|\rho^{PT}|$ are defined over the interval $|\rho^{PT}| \in [-\frac{1}{16},\frac{1}{256}]$, while the associated separability probabilities are the cumulative probabilities of these distributions over the nonnegative 
subinterval $|\rho^{PT}| \in [0,\frac{1}{256}]$. We note that for the fully mixed (classical) state, $|\rho^{PT}| = \frac{1}{256}$, while for a maximally entangled state, such as  a Bell state,  $|\rho^{PT}| = -\frac{1}{16}$.)

A highly-intriguing aspect of the (not yet rigorously established) determinantal moment formulas obtained (by C. Dunkl) in \cite[App.D.4]{MomentBased} was that both the two-rebit and two-qubit cases could be encompassed by a 
{\it single} formula, with a Dyson-index-like parameter 
$\alpha$ \cite{MatrixModels} serving to distinguish the two cases. The value $\alpha = \frac{1}{2}$ corresponded to the two-rebit case and $\alpha=1$ to the two-qubit case. (Let us note that the results of the formula for $\alpha=2$ and $n=1$ and 2 have recently been confirmed computationally by Dunkl using the "Moore determinant" (quasideterminant) \cite{Moore,Gelfand} of $4 \times 4$ quaternionic density matrices. However, tentative efforts of ours to verify the $\alpha=4$ 
[conjecturally, {\it octonionic} 
\cite{LiaoWangLi}, problematical] case, have not proved successful.)

When the probability-distribution-reconstruction algorithm \cite{Provost} was applied 
in \cite{MomentBased} to the two-rebit case ($\alpha=\frac{1}{2}$), employing the first 3,310 moments of $|\rho^{PT}|$, 
a (lower-bound) estimate that was 0.999955 times as large as $\frac{29}{64} \approx 0.453120$ was obtained (cf. \cite[p. 6]{advances}). Analogously, in the two-qubit case ($\alpha =1$), using 2,415 moments, an estimate that was 0.999997066 times as large as $\frac{8}{33} \approx 0.242424$ was derived. This constitutes an appealingly simple rational value that had previously been conjectured in a quite different (non-moment-based) form of analysis, in which "separability functions" had been the main tool employed \cite{slater833}.

Further, the determinantal moment formulas advanced in 
\cite{MomentBased} were then applied with $\alpha$ set equal to 2. This appears--as the indicated recent (Moore determinant) computations of Dunkl 
show--to correspond to the generic 27-dimensional set of quaternionic density matrices \cite{andai,adler}. Quite remarkably, a separability probability estimate, based on 2,325 moments, that was 0.999999987 times as large as $\frac{26}{323} \approx 0.0804954$ was found. (In line with this set of three results, the paper \cite{MomentBased} was entitled, "Moment-Based Evidence for Simple Rational-Valued Hilbert-Schmidt Generic $2 \times 2$ Separability Probabilities".)

In the present study, we extend these three (individually-conducted) moment-based analyses in a more systematic, thorough manner, {\it jointly} embracing the sixty-four integral and half-integral values $\alpha =\frac{1}{2}, 1, \frac{3}{2}, 2,\ldots, 32$. We do this by accelerating, for our specific purposes, the Mathematica probability-distribution-reconstruction program of Provost \cite{Provost},  in a number of ways. Most significantly, we make use of the three-term recurrence relations for the Legendre polynomials. Doing so obviates the need to compute each successive higher-degree Legendre polynomial {\it ab initio}.

\begin{figure}
\includegraphics{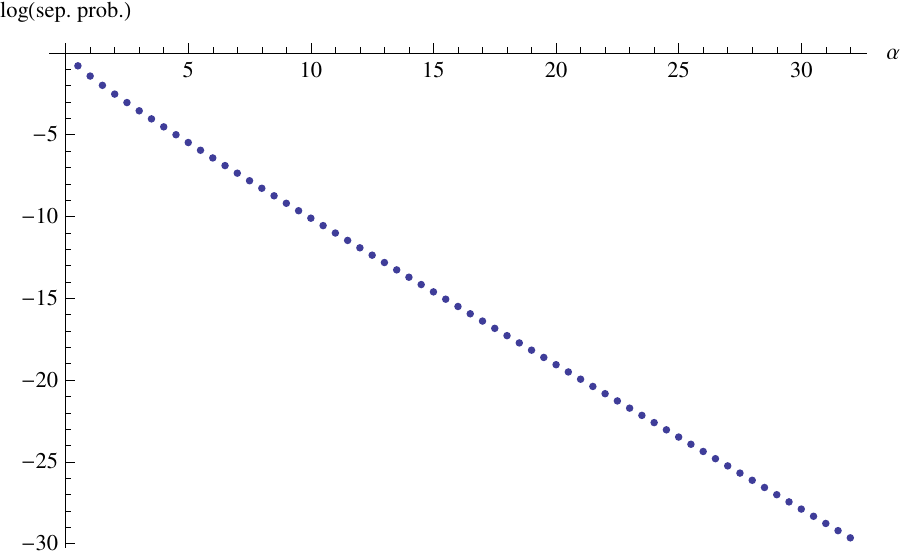}
\caption{\label{fig:ListPlotLogEstimates}Logarithms of generalized separability probability estimates, based on 7,501 Hilbert-Schmidt moments of $|\rho^{PT}|$, as a function of the Dyson-index-like parameter $\alpha$}
\end{figure}

\begin{figure}
\includegraphics{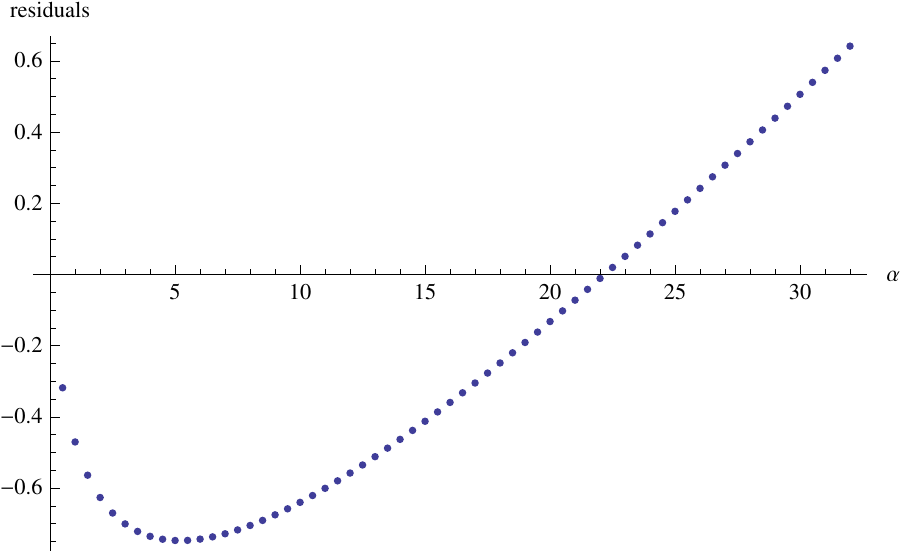}
\caption{\label{fig:EnlargedResiduals}Residuals from linear fit to logarithms of generalized separability probability estimates}
\end{figure}

In this manner, we were able to obtain--using exact computer arithmetic
throughout--"generalized" separability probability estimates based on 7,501 moments for $\alpha = \frac{1}{2}, 1, \frac{3}{2},\ldots,32$.
In Fig.~\ref{fig:ListPlotLogEstimates} we plot the logarithms of the resultant sixty-four separability probability estimates (cf. \cite[Fig. 8]{MomentBased}), which fall close to the line $-0.9464181889 \alpha$.
In Fig.~\ref{fig:EnlargedResiduals} we show the residuals from this linear fit.

Most notably, in Fig.~\ref{fig:HypergeometricFormula1} we will present a hypergeometric-function-based formula, together with striking supporting evidence
for it, that appears to succeed in uncovering the functional relation
($P(\alpha)$) 
underlying these generalized separability probabilities. Further, in 
(\ref{strikingresults}), and the immediately preceding text, 
we list a number of remarkable values yielded by the hypergeometric 
formula for values of $\alpha$ other than the basic 64 (half-integral and integral) values from which we have started.

\section{Results}
\subsection{The three basic conjectures revisited} \label{threebasics}
\subsubsection{$\alpha=\frac{1}{2}$--the two-rebit case}
In \cite{MomentBased}, a lower-bound estimate of the two-rebit separability probability was obtained, with the use of the first 3,310 moments of $|\rho^{PT}|$. It was 0.999955 times as large as
$\frac{29}{64} \approx 0.453120$. With the use, now, of 7,501 moments, the figure increases to 
0.999989567. This outcome, thus, fortifies our previous conjecture.

\subsubsection{$\alpha=1$--the two-qubit case}
In \cite{MomentBased}, a lower-bound  estimate of the two-qubit separability probability was obtained, with the use of the first 2,415 moments of $|\rho^{PT}|$, that was 0.999997066 times as large as
$\frac{8}{33} \approx 0.242424$. Employing 7,501 moments, this figure increases to 0.99999986.

\subsubsection{$\alpha=2$--the  quaternionic case}
In \cite{MomentBased}, a lower-bound  estimate of the quaternionic separability probability was obtained that was 0.999999987 times as large as $\frac{26}{323} \approx 0.0804954$, using the first 2,325 moments of $|\rho^{PT}|$. Based on 7,501 moments, this figure increases, quite remarkably still, to 0.999999999936.

\subsection{Generalized separability probability formula}

A principal motivation in undertaking the analyses reported here--in addition, to further scrutinizing the three specific conjectures reported in \cite{MomentBased}--was to uncover the functional relation
underlying the curve in Fig.~\ref{fig:ListPlotLogEstimates} (and/or its original non-logarithmic counterpart).

Preliminarily, let us note that the {\it zeroth}-order approximation (being independent of the particular value of $\alpha$) provided by the Provost probability-distribution-reconstruction algorithm is simply the {\it uniform} distribution over the interval
$[-\frac{1}{16},\frac{1}{256}]$. The corresponding zeroth-order separability probability estimate is the cumulative probability of this distribution over the nonnegative subinterval $[0,\frac{1}{256}]$, that is, $ \frac{1}{256}/(\frac{1}{16} +\frac{1}{256}) =\frac{1}{17} \approx 0.0588235$. So, it certainly appears that speedier convergence (sec.~\ref{threebasics}) of the algorithm occurs for separability probabilities, the true values of which are initially close to 
$\frac{1}{17}$ (such as $\frac{26}{323} \approx 0.0804954$ 
in the quaternionic case).
Convergence also markedly increases as $\alpha$ increases.

It appeared, numerically, that the generalized separability probabilities for integral and half-integral values of $\alpha$ were rational values 
(not only for the three specific values $\alpha = \frac{1}{2}, 1, 2$ of original focus). With various computational tools and search strategies based upon emerging mathematical properties, we were able to advance additional,  seemingly plausible conjectures as to the exact values for $\alpha=3, 4, \ldots,32$, as well.
(We inserted many of our high-precision numerical estimates 
into the search box
on the Wolfram Alpha website--which then indicated likely candidates for corresponding rational values.) 

We fed this sequence of thirty-two conjectured rational numbers into the FindSequenceFunction command of Mathematica. (This command "attempts to find a simple function that yields the sequence $a_i$ when given successive integer arguments.") To our considerable satisfaction, this produced a generating formula (incorporating a diversity of hypergeometric functions of the $_{p}F_{p-1}$ type, $p=7,\ldots,11$, {\it all} with argument $z =\frac{27}{64}= (\frac{3}{4})^3$) for the sequence. (Let us note that $z^{-\frac{1}{2}} = \sqrt{\frac{64}{27}}$ is the "residual entropy for square ice" \cite[p. 412]{finch} (cf. \cite[eqs.[(27), (28)]{ckksr}) \cite{guillera}. In fact, the Mathematica command succeeds using only the first twenty-eight conjectured rational numbers, but no fewer.)

However, the formula produced was quite cumbersome in nature (extending over several pages of output). With its use, nevertheless, we were able to convincingly
generate rational values for {\it half}-integral $\alpha$ (including the two-rebit $\frac{29}{64}$ conjecture), also fitting our corresponding half-integral thirty-two numerical estimates exceedingly well.
(Let us strongly emphasize that the hypergeometric-based formula 
was generated using {\it only} the integral values of $\alpha$. The process was fully reversible, and we could first employ the half-integral results
to generate the formula--which then--seemingly perfectly fitted the integral values.)

At this point, for illustrative purposes, let us list the first ten half-integral and ten integral rational values (generalized separability probabilities), along with their approximate numerical values.

\begin{equation}
\begin{array}{cc}
 
\begin{array}{cccc}
 \text{$\alpha $ =} & \frac{1}{2} & \frac{29}{64} & 0.453125 \\
\end{array}
 & 
\begin{array}{cccc}
 \text{$\alpha $ =} & 1 & \frac{8}{33} & 0.242424 \\
\end{array}
 \\
 
\begin{array}{cccc}
 \text{$\alpha $ =} & \frac{3}{2} & \frac{36061}{262144} & 0.137562 \\
\end{array}
 & 
\begin{array}{cccc}
 \text{$\alpha $ =} & 2 & \frac{26}{323} & 0.0804954 \\
\end{array}
 \\
 
\begin{array}{cccc}
 \text{$\alpha $ =} & \frac{5}{2} & \frac{51548569}{1073741824} & 0.0480083 \\
\end{array}
 & 
\begin{array}{cccc}
 \text{$\alpha $ =} & 3 & \frac{2999}{103385} & 0.0290081 \\
\end{array}
 \\
 
\begin{array}{cccc}
 \text{$\alpha $ =} & \frac{7}{2} & \frac{38911229297}{2199023255552} & 0.0176948 \\
\end{array}
 & 
\begin{array}{cccc}
 \text{$\alpha $ =} & 4 & \frac{44482}{4091349} & 0.0108722 \\
\end{array}
 \\
 
\begin{array}{cccc}
 \text{$\alpha $ =} & \frac{9}{2} & \frac{60515043681347}{9007199254740992} &
   0.00671852 \\
\end{array}
 & 
\begin{array}{cccc}
 \text{$\alpha $ =} & 5 & \frac{89514}{21460999} & 0.00417101 \\
\end{array}
 \\
 
\begin{array}{cccc}
 \text{$\alpha $ =} & \frac{11}{2} & \frac{71925602948804923}{27670116110564327424} &
   0.0025994 \\
\end{array}
 & 
\begin{array}{cccc}
 \text{$\alpha $ =} & 6 & \frac{179808469}{110638410169} & 0.00162519 \\
\end{array}
 \\
 
\begin{array}{cccc}
 \text{$\alpha $ =} & \frac{13}{2} &
   \frac{3387374833367307236269}{3324546003940230230441984} & 0.0010189 \\
\end{array}
 & 
\begin{array}{cccc}
 \text{$\alpha $ =} & 7 & \frac{191151001}{298529164591} & 0.000640309 \\
\end{array}
 \\
 
\begin{array}{cccc}
 \text{$\alpha $ =} & \frac{15}{2} &
   \frac{124792688228667229196729}{309485009821345068724781056} & 0.000403227 \\
\end{array}
 & 
\begin{array}{cccc}
 \text{$\alpha $ =} & 8 & \frac{1331199762}{5232880523393} & 0.000254391 \\
\end{array}
 \\
 
\begin{array}{cccc}
 \text{$\alpha $ =} & \frac{17}{2} &
   \frac{407557367133399293946182513}{2535301200456458802993406410752} & 0.000160753 \\
\end{array}
 & 
\begin{array}{cccc}
 \text{$\alpha $ =} & 9 & \frac{74195568677}{729345064647247} & 0.000101729 \\
\end{array}
 \\
 
\begin{array}{cccc}
 \text{$\alpha $ =} & \frac{19}{2} &
   \frac{1338799759394288468677657208071}{20769187434139310514121985316880384} &
   0.0000644609 \\
\end{array}
 & 
\begin{array}{cccc}
 \text{$\alpha $ =} & 10 & \frac{730710456538}{17868447453498669} & 0.0000408939 \\
\end{array}
 \\
\end{array}
\end{equation}

\begin{figure}
\includegraphics{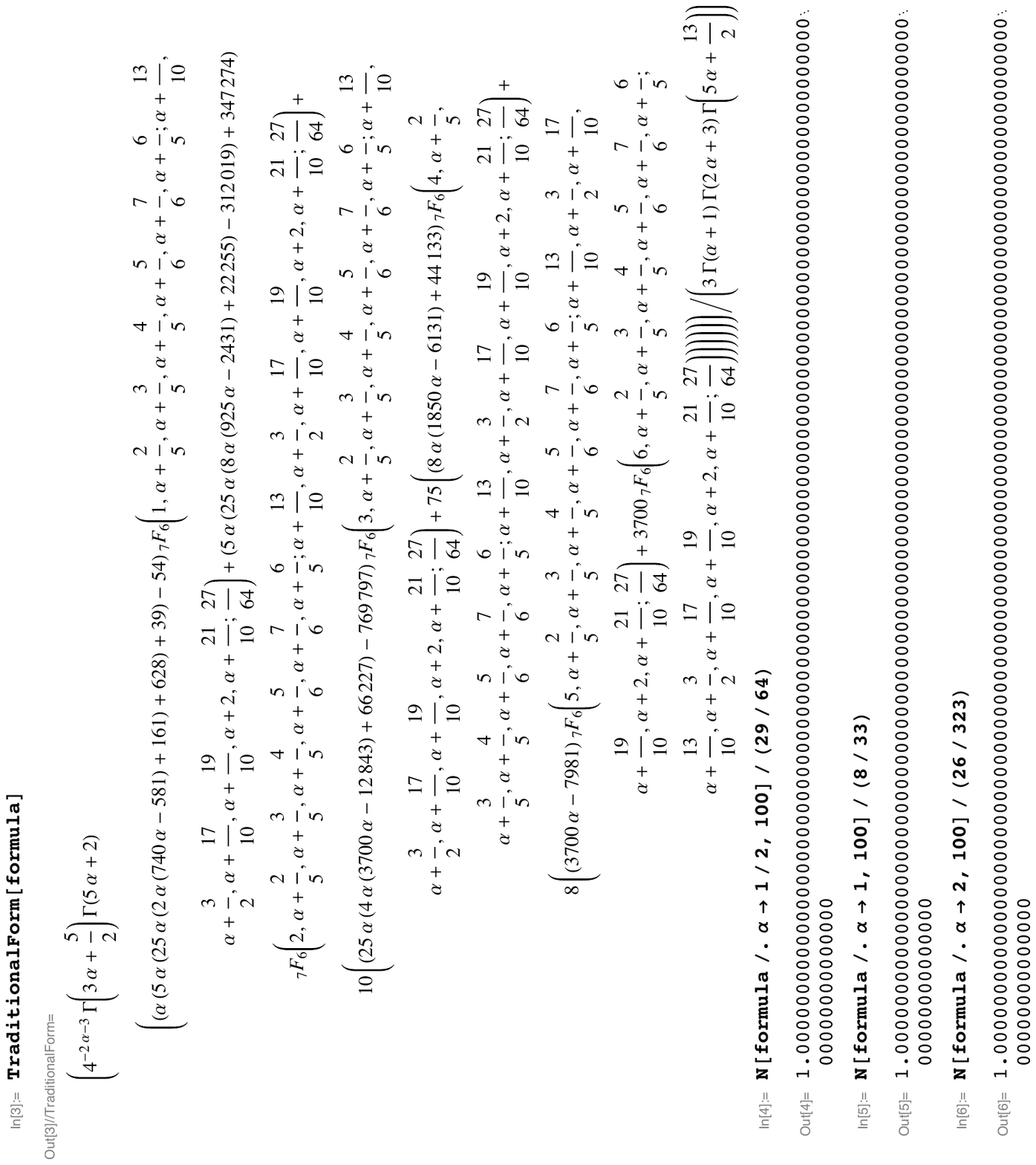}
\caption{\label{fig:HypergeometricFormula1}Hypergeometric formula $P(\alpha)$ for Hilbert-Schmidt generic $2 \times 2$ {\it generalized} separability probabilities and evidence that it reproduces the basic three (real [$\alpha = \frac{1}{2}$], complex [$\alpha = 1$] and quaternionic [$\alpha = 2$]) conjectures of 
$\frac{29}{64}, \frac{8}{33}$ and $\frac{26}{323}$}
\end{figure}

To simplify the cumbersome (several-page) output yielded by the Mathematica FindSequenceFunction command, we employed certain of the "contiguous rules" for hypergeometric functions listed by 
C. Krattenthaler in his package HYP \cite{ck} (cf. \cite{bytev}). Multiple applications of the rules C14 and C18 there, together with certain gamma function simplifications suggested by C. Dunkl, led to the rather more compact formula displayed in 
Fig.~\ref{fig:HypergeometricFormula1}. (Attempts to achieve a still more succinct form  
have not yet succeeded.)
This formula  incorporates a six-member family  ($k =1,\ldots,6$) of hypergeometric functions, differing only in the first upper index $k$,
\begin{equation} \label{family}
\, _7F_6\left(k,\alpha +\frac{2}{5},\alpha +\frac{3}{5},\alpha +\frac{4}{5},\alpha
   +\frac{5}{6},\alpha +\frac{7}{6},\alpha +\frac{6}{5};\alpha +\frac{13}{10},\alpha
   +\frac{3}{2},\alpha +\frac{17}{10},\alpha +\frac{19}{10},\alpha +2,\alpha
   +\frac{21}{10};\frac{27}{64}\right) .
\end{equation}
Interestingly, we are only able to, in general, evaluate the formula numerically, but then to arbitrarily high (hundreds, if not thousand-digit) precision, giving us strong confidence in the validity of the {\it exact} generalized separability probabilities that we advance. Whether exact values can be directly extracted, without recourse to such high-precision numerics, appears to be an issue still yet to be fully resolved.

Let us now apply the formula (Fig.~\ref{fig:HypergeometricFormula1}) to values of $\alpha$ other than the basic sixty-four. For $\alpha = 0$, the formula yields--as would be expected--the "classical separability probability" of 1. Further, proceeding in a purely formal manner (since there appears to be no corresponding  genuine probability distribution over $[-\frac{1}{16},\frac{1}{256}]$), for the {\it negative} value $\alpha = - \frac{1}{2}$, the formula  yields $\frac{2}{3}$. For $\alpha =-\frac{1}{4}$, it gives -2. Remarkably still, for $\alpha = \frac{1}{4}$, the result is clearly (to one thousand decimal places) equal to $2-\frac{34}{21 \text{agm}\left(1,\sqrt{2}\right)} =  2-\frac{17 \Gamma \left(\frac{1}{4}\right)^2}{21 \sqrt{2} \pi ^{3/2}} \approx 
0.6486993992$, where the arithmetic-geometric mean of 1 and $\sqrt{2}$ is indicated. (The reciprocal of this mean is Gauss's constant.) For $\alpha = \frac{3}{4}$, the result equals 
$2-\frac{9689 \Gamma \left(\frac{3}{4}\right)}{4420 \sqrt{\pi } \Gamma
   \left(\frac{5}{4}\right)} \approx 0.3279684732$, while for $\alpha=-\frac{3}{4}$, we have
$\frac{128}{21 \text{agm}\left(1,\sqrt{2}\right)}+2 =2+\frac{32 \sqrt{2} \Gamma \left(\frac{1}{4}\right)^2}{21 \pi ^{3/2}} \approx 7.087249321$.  For $\alpha=\frac{2}{3}$, the outcome is $2-\frac{288927 \Gamma \left(\frac{1}{3}\right)^3}{344080 \pi ^2} \approx 0.36424897456$.
Results are presented in the table 
\begin{equation} \label{strikingresults}
\left(
\begin{array}{ccc}
 \alpha  & P(\alpha ) & \text{value} \\
 -\frac{3}{4} & 2+\frac{32 \sqrt{2} \Gamma \left(\frac{1}{4}\right)^2}{21 \pi ^{3/2}} &
   7.08725 \\
 -\frac{2}{3} & 2-\frac{8 \pi }{\sqrt{3} \Gamma \left(\frac{1}{3}\right)^3} & 1.24527
   \\
 -\frac{1}{2} & \frac{2}{3} & 0.666667 \\
 -\frac{1}{3} & 2+\frac{3 \Gamma \left(\frac{1}{3}\right)^3}{4 \pi ^2} & 3.461 \\
 -\frac{1}{4} & 2 & 2 \\
 \frac{1}{4} & 2-\frac{17 \Gamma \left(\frac{1}{4}\right)^2}{21 \sqrt{2} \pi ^{3/2}} &
   0.648699 \\
 \frac{1}{3} & 2-\frac{459 \sqrt{3} \pi }{91 \Gamma \left(\frac{1}{3}\right)^3} &
   0.572443 \\
 \frac{2}{3} & 2-\frac{288927 \Gamma \left(\frac{1}{3}\right)^3}{344080 \pi ^2} &
   0.364249 \\
 \frac{3}{4} & 2-\frac{9689 \Gamma \left(\frac{3}{4}\right)}{4420 \sqrt{\pi } \Gamma
   \left(\frac{5}{4}\right)} & 0.327968 \\
\end{array}
\right) .
\end{equation}
(Let us note that the term 
$\frac{3 \Gamma \left(\frac{1}{3}\right)^3}{4 \pi ^2} \approx 1.46099848$ present in the result for 
$\alpha =-\frac{1}{3}$ is "Baxter's four-coloring constant" for a triangular lattice \cite[p. 413]{finch}.) Also, for $\alpha=-1$, we have $\frac{2}{5}$. For $\alpha=-\frac{3}{2}$, the result is $-\frac{2}{3}$.

Thus, it appears that our hypergeometric-related formula 
(Fig.~\ref{fig:HypergeometricFormula1}) constitutes a major advance in addressing the long-standing (Hilbert-Schmidt) $2 \times 2$ separability probability question \cite{ZHSL}. However, there certainly remain the important problems of formally verifying this formula (as well as the underlying determinantal moment formulas in \cite{MomentBased}, employed in the probability-distribution reconstruction process), and achieving a better understanding of what it conveys regarding the geometry of quantum states \cite{ingemarkarol}. Further, questions of the asymptotic behavior of the formula ($\alpha \rightarrow \infty$), possible modified/simplified/canonical forms of it, and of possible Bures metric 
\cite{szBures,ingemarkarol,slaterJGP,slaterqip,slaterC} counterparts, are under current investigation.

\begin{acknowledgments}
I would like to express appreciation to the Kavli Institute for Theoretical
Physics (KITP)
for computational support in this research, and to Christian Krattenthaler, Charles F. Dunkl and Michael Trott for their expert 
advice. K. {\.Z}yczkowski--as always--provided encouragement.
\end{acknowledgments}

\bibliography{HypergeometricFormula2}

\end{document}